
\magnification=\magstep1
\newcount\startpage
\startpage=1
\pageno=\startpage
\def\makeheadline{\vbox to 0pt{\vskip -40.0 truept
    \line{\vbox to8.5pt{}\the\headline}\vss}\nointerlineskip}
\hsize=16.35 truecm
\vsize=23.3 truecm
\footline{\hfill}
\def\leftheadline{\tenrm\folio\hss {\it Klypin \& Rhee}}
\def\rightheadline{\it {The cluster-cluster correlation
function}\hss\tenrm\folio}
\def \firsthead{\hss\tenrm\folio}
\headline{\ifnum\pageno=\startpage \firsthead \fi \ifnum\pageno>\startpage
\ifodd\pageno\rightheadline\else\leftheadline\fi\fi}
\def\deg{\ifmmode^\circ\else$^\circ$\fi\ }

\def\solar{\ifmmode_{\mathord\odot}\else$_{\mathord\odot}$\fi}
\def\h1m{h^{-1}}
\def\h1{$h^{-1}$}
\def \Msun{{\rm M}_{\odot}}

\def\arcm{\char'023\ }
\def\kms{~km~s$^{-1}$\ }

\def\etal{et~al.\ }
\def\etalc{et~al.,\ }
\def\Mpch {h^{-1}{\rm Mpc}}
\def\Mpc {{\rm Mpc}}
\def \r0 {r_0}
\def \rcc {r_{cc}}
\def\solar{\ifmmode_{\mathord\odot}\else$_{\mathord\odot}$\fi }
\null
\def\kms{~km~s$^{-1}$\ }
\def\arcm{\ifmmode {' }\else $' $\fi}
\magnification=\magstep1\vsize=20cm
\baselineskip=1truecm
\centerline{\bf THE ZERO-POINT OF THE CLUSTER-CLUSTER CORRELATION
FUNCTION:}
\centerline{\bf A KEY TEST OF COSMOLOGICAL POWER SPECTRA}
\vskip 2.0 truecm
\centerline{ANATOLY KLYPIN$^1$ and GEORGE RHEE$^{1,2}$}
\medskip
\centerline{$^1$ Department of Astronomy}
\centerline{New Mexico State University}
\centerline{Las Cruces, NM 88003-0001}
\medskip
\centerline{$^2$ Department of Physics}
\centerline{University of Nevada Las Vegas}
\centerline{Las Vegas, NV 89154-4002}
\bigskip
\bigskip
\bigskip
\vskip 5 truecm
\centerline {Received \underbar{\hskip 3truecm}}
\vfill\eject
\vskip 5truecm
\centerline {\bf ABSTRACT}
\medskip\noindent
We propose the zero-point of the cluster-cluster correlation function
as a sensitive test for the shape of the power spectrum of initial
fluctuations.
With a wealth of new available redshifts for rich clusters
the correlation function is measured to higher
accuracy than has previously been done. In particular it is now possible
to go beyond the power law description to measure the point at which
the
correlation function becomes zero. Four independent
measurements of the zero-point of the rich galaxy cluster correlation
function indicate that the zero point, $r_0$, should be in the range
(40-60)$h^{-1}$Mpc.
The large value of $r_0$ at which the zero-point occurs rules out
conventional
CDM models independently of the assumed amplitude.
The most severe constraints are imposed on the CDM models with the
cosmological constant. Models with $\Omega < 0.25$ should be rejected
because they predict too large $r_0$. If the age of the Universe
is assumed to be larger than 15 Gyr, models with either $\Omega<0.5$ or
$h> 0.55$ are rejected.

We present the results of numerical simulations of clusters in
an $\Omega=1$ cosmological model with a mixture of cold plus hot dark
matter (CHDM).
We have identified clusters in the simulations and calculated the
cluster-cluster correlation function. The function we determined
for the simulated clusters has a zero-point, $r_0=55h^{-1}$Mpc
that accurately matches
the zero point of the observed function. In addition the shape
of the function on all available scales (up to $100h^{-1}$Mpc) is
reproduced for both the Abell and the APM clusters.

\smallskip
\noindent {\it Subject headings:} cosmology: theory ---
dark matter --- large scale structure of the universe ---
galaxy clusters
\vfill\eject
\centerline{\bf 1. INTRODUCTION}
\smallskip\noindent
The clustering properties of rich clusters of galaxies can provide a
powerful
constraint for theories of galaxy formation. In particular the cluster
spatial two-point correlation function provides a useful quantitative
measure of clustering for comparison with theories. The first
determination
of the cluster correlation function for a complete redshift sample
gave $\xi_{cc}=(r/r_{cc})^{-1.8}$
where $r_{cc}=25\Mpch$ (Bahcall and Soneira, 1983, Klypin and Kopylov,
1983). The correlation amplitude is much larger than that of galaxies
$r_{gg} \sim 5 \Mpch$ (Davis and Peebles, 1983).

Since the first work on the cluster correlation function there has
been
controversy concerning the true value of the correlation function
amplitude (i.e. $r_{cc}$). Sutherland (1988) and Sutherland and
Efstathiou (1991)
have claimed that the selection of Abell clusters is non-uniform on
the sky
with a tendency to find pairs of clusters close in projection but at
very different redshifts. These authors claim a lower correlation than
that found by Bahcall and Soneira (1983); $\rcc=14\Mpch$. Other studies
made on subsamples which not suffer from the biases claimed by
Sutherland for the whole Abell catalog found an intermediate value
for the correlation length $\rcc=20\Mpch$. These studies include
cD clusters (West and van den Bergh, 1990) and X-ray clusters
(Lahav \etal 1989).  Most recently the studies of Peacock and West
(1992) and Postman \etal (1992), using samples of 427 and 351 clusters
respectively, find a correlation length close to $20 \Mpch$.

To settle the issue of the reliability of Abell's cluster catalog
automated cluster surveys have been made. Two recent projects are
the Automated Plate Measuring Facility (APM) survey (Dalton \etal 1992)
and the Edinburgh-Durham Cluster Catalog (Nichol \etal 1992). The
correlation functions for clusters found in the catalogs have slopes
similar to previous ones and amplitudes ($\sim 15 \Mpch$) lower than
previously determined for the Abell catalog.

On the theoretical front one can use models for the growth of structure
in the universe by gravitational instability to make predictions
concerning the amplitude and shape of the cluster correlation
function. These predictions will vary depending on the initial
fluctuation spectrum which in turn is determined by the physics
of the dark matter particles.
Bahcall and Cen (1992) have found that
standard biased cold dark matter models ($\Omega=1$) are inconsistent
with
the observed cluster mass function and cluster correlation function
for any bias parameter. A low density,
low-bias ($\Omega \sim 0.25$, $b \sim 1$)
CDM model with or without a cosmological constant appears to be consistent
with both the cluster correlation length, $\rcc$, and the cluster mass
function. Jing \etal, (1993) have calculated two point correlation
functions of clusters in a set of CHDM models.
They
found that a hybrid model of the universe which contains (in mass)
$\sim 30\%$ HDM, $\sim 70\%$ CDM and $\geq 1\%$ baryons could
provide a reasonable fit to the observed two point correlation
function of Abell clusters. Note however that they were mainly
concerned with $\xi_{cc}(r)$ for $r<50 \Mpch$ and did not
actually simulated the hot component.
Olivier \etal (1993) have compared
the cluster correlation function with the CDM and Textures models.
They find that for 40 Mpc $ \leq r \leq$ 80 Mpc the observed
correlation
function is larger than that predicted by these two theories. Holtzman
and Primack (1993) using the peaks formalism for gaussian density
fields
have investigated the prediction for the cluster
correlation function in high and low $\Omega$ CDM universes and in the
C+HDM model. They find that the data are best fit by the CHDM model.
Bartlett and Silk (1993) have reached the same conclusion by making
analytical model predictions for the cluster X-ray temperature
function, based on the Press-Schechter approximation.

Now that cluster catalogs containing redshifts for several hundred
clusters have become available it is possible to determine the
scale at which the cluster correlation function has a value of
zero (we refer to this as the $\xi_{cc}$ zero-point). We argue in this
paper that this is a key observation for comparison with models.
We know of four independent determinations of the correlation
function zero-point (Scaramella \etal 1993, Dalton \etal 1992, Postman
\etal 1992 and Postman 1993,
 Peacock and West 1992).

\bigskip
\centerline{\bf 2. COSMOLOGICAL MODELS}
\noindent
We study the CHDM model, which assumes that
the Universe has the critical density and the Hubble parameter is
$H=100h$\kms\ Mpc$^{-1}$, $h=0.5$ and the cosmological constant is
zero.
The CHDM model has the following parameters. The mass density of the
Universe in the form of neurtrino is $\Omega_{\nu}=0.30$, density of
baryons is $\Omega_b=0.1$.
We use analytical approximations (c.f. eq.(1) of Klypin \etalc 1993)
for the ``cold'' and ``hot'' spectra.
The amplitude of fluctuations is normalized so that our
realization is drawn from an ensemble producing the
quadrupole in the angular fluctuations of the cosmic
microwave background at the $17\mu K$ level measured by COBE.
This gives the rms fluctuations of mass in a sphere of
$8h^{-1}$Mpc radius, $\sigma_8=0.667$ and the rms velocity of the
``cold'' dark matter relative to the rest frame $\sigma_v=750$ \kms.

\bigskip
\centerline{\bf 3. N-BODY SIMULATIONS}
\noindent
Numerical simulations are done using a standard Particle-Mesh
(PM) code (Hockney \& Eastwood 1981, Kates, Kotok, \& Klypin
1991). Two runs with different realizations of the initial spectrum
have been done. Both runs are simulated  using a $256^3$ mesh
for the gravitational force resolution.
Each simulation has $128^3$ ``cold'' particles and two additional sets
of $128^3$ particles in each set to represent ``hot''
neutrinos. We use the same prescription to simulate random thermal
velocities of ``hot'' particles as Klypin \etal (1993). ``Hot''
particles are generated in pairs, particles of each
pair having random ``thermal'' velocities of exactly equal
magnitude but pointing in opposite directions. The directions
of these ``thermal'' velocities are random. The magnitudes
of the velocities are drawn from relativistic Fermi-Dirac
statistics.  The particles have different relative
masses: each ``cold'' particle has a relative mass 0.7 and
each ``hot'' particle has the mass $0.3/2=0.15$.

Initial
positions and velocities of particles were set using the
Zel'dovich approximation.  The displacement
vector was simulated directly. Phases of fluctuations were
exactly the same for ``hot'' and ``cold'' particles. When
generating velocities of ``hot'' particles, the  ``thermal''
component, as described above, was added to the velocity
produced by the Zel'dovich approximation.

The size of the computational box
for the simulations is 400Mpc ($h=0.5$). The
simulations were started at redshift $z=7$ and were run to redshift
zero with a constant step $\Delta a=0.01$ in the expansion
parameter $a$.
The smallest resolved comoving scale  in these simulations is
$0.78\Mpch$  and the mass of a ``cold''
particle was  $7.33\times 10^{11}h^{-1}M\solar$.

\bigskip
\centerline{\bf 4. IDENTIFICATION OF GALAXY CLUSTERS IN THE
SIMULATIONS}

Observationally real galaxy clusters are found as local concentrations
within some radius. For the Abell clusters the radius
is $1.5\Mpch$ and for APM clusters it is
$0.75\Mpch$. We mimic this procedure for our simulations by finding
all local  maxima of  the total (``cold'' plus ``hot'')
density above density threshold $\delta\rho/\rho > 100$. The density
field is produced on the $256^3$  grid, which corresponds to the
spacial resolution $0.78\Mpch$. The effective smoothing scale,
corresponding to this mesh is smaller
than that of real galaxy clusters. As a result, the number of
selected maxima
is much larger than the expected number of galaxy clusters for
this volume (about 100). At $z=0$ the number of maxima was 1059 in one
simulation and 987 in another. Since the scale is
not very far from the radius of real clusters, we use the density
maxima as a starting point for the search of the clusters.
A sphere of cluster radius is placed at each density maximum and
the centrum of mass of the sphere is found. Then the centre of the
sphere is displaced to the centrum of mass and the procedure is
repeated. As the result the spheres move in space in the
process of searching for maximum of the number of dark matter
particles inside the sphere radius. After five iteration the process
converges and positions of spheres stop to change. For one of the
simulations at $z=0$ we made 10 iterations and found that there was no
difference from results obtained after 5 iterations. Some of the
spheres (about 10 per cent) find the same clusters. In order to
remove the duplicates, pairs of clusters with distance between centers
less than  $0.75\Mpch$ are found and only one more massive cluster
in a pair is retained for the subsequent analysis. We still have more
candidates than the expected number of rich galaxy clusters in our
volume because most of our ``clusters'' are less massive than Abell or
APM clusters.

\bigskip
\centerline{\bf 5. CLUSTER PROPERTIES IN THE CHDM MODEL}
\smallskip
\noindent
For clusters found in the APM survey Dalton \etal (1992) give
$n=2.4\times 10^{-5}h^3$Mpc$^{-3}$
for the richness class ${\cal R} > 20$.
This would mean 195 clusters in our volume. By imposing the mass
threshold $M> 1.05\times 10^{14}h^{-1}\Msun$ on our APM-style
``clusters'', we find 203 clusters in one simulation and 206 in
another, which is close to the expected number of APM clusters.
The RMS 1D peculiar velocity of the clusters is $\sigma_v=418$\kms.
This is
compatible with the estimate given by Dalton \etal (1992): they
formally estimate $\sigma_v=700$\kms, but measuring errors account for
about half of the dispersion. This brings the estimate of real
velocities to the level of 500\kms, with a large uncertainty.
However, not all of the rms velocity 418\kms in the model
take part
in producing anisotropy of $\xi_{\rm cc}(\sigma,\pi)$, which was the
method used by Dalton \etal to estimate to cluster velocities.
Most of it is due to very long waves. As a rough estimate, we should
remove from the estimate the velocities of a sphere of radius
$40\Mpch$, because the anisotropy of $\xi_{\rm cc}(\sigma,\pi)$ was
found well within the radius. This reduces the velocity to the level
$\sigma_v =(300-350)$\kms. It is not clear if this estimate contradicts
results of Dalton \etal because the measuring errors are too close to
the estimated $\sigma_v$. Dalton \etal also give the number of
clusters with richness ${\cal R} > 35$. If one assumes that the mass of
a cluster is proportional to its richness, than the CHDM model
produces a factor of two smaller of APM clusters of
richness ${\cal R} > 35$:
there are  45 and 46 clusters in our simulations while the observed
number is 94. The difference could be due to numerical effects. If
more massive APM-style cluster is also slightly thicker and its
diameter is larger than 2 cells, we underestimate its mass.

Figure 1 shows the number density of Abell clusters in the CHDM model
as a function of redshift (full curves and filled circles). The
Press-Schechter
approximation does not give a good fit in this case. If it is fitted to
the mass function at $z=0, \delta_c=1.50$, it underestimates by a
factor of 3 the
number of clusters with mass larger than $2\times 10^{14}h^{-1}\Mpc$
at $z=0.5$. The dashed curves show the PS results for $\delta_c=1.45$.

The full curve in Figure~2 shows the mass
function of Abell clusters in the CHDM model. In order to give a rough
estimate of the errors we also show the
mass function in the two simulations separately as markers. The
dot-dashed curve is for the mass function obtained by the
Press-Schechter approximation with the parameter $\delta_c =1.50$.
The dashed curve shows  the mass function of Abell clusters estimated
by  Bahcall \& Cen (1993).  The errors associated with their results
are probably larger than formal statistical
uncertainties given by Bahcall \& Cen (1993).
However, the fit given by Bahcall \& Cen is not very far from the
results obtained in our simulations. The upper dashed curve provides a
very accurate fit to the mass function in the CHDM model. It is the
same fit as found by Bahcall \& Cen, but the mass scale is higher by
the factor 1.6. Thus, the mass function of Abell clusters in the CHDM
is well approximated by
$$
  n(>M) = 4\times 10^{-5}(M/M_*)^{-1}\exp(-M/M_*)h^3{\rm Mpc}^{-3},
           \quad M_*=2.9\times 10^{4}h^{-1}M\solar~.
$$

There are some
possible reasons for the discrepancy with results of Bahcall \&
Cen. One of them is due to the fact that a large fraction of cluster
mass is gained at relatively
large radii from the center of the cluster center. Thus, the total
mass significantly depends on such assumptions as the density profile
or sphericity. (Note also that ``significant'' in this context means 30--50
percent). Note that an independent study by Biviano \etalc (1993)
yields data which lie slightly {\it above} our model curve as opposed
to Bahcall \& Cen data which lie below it.
Note also that uncertainties in the estimates for the total mass of the
Coma cluster are sufficient to account
for this factor. For example, for the mass of the Coma cluster,
White \etal (1993) give $1.1\cdot 10^{15}h^{-1}\Msun$ as compared with
$0.65\cdot 10^{15}h^{-1}\Msun$ assumed by Bahcall \& Cen. It seems
that the systematic errors dominate the statistical ones in estimating
the observed cluster mass function.

Another way to characterize the mass distribution of clusters is
to show the x-ray temperature function of the clusters, $dn/dT$, which
is the number density of clusters with given hot gas temperature. The
temperature of the gas in clusters is less sensitive to details of
the density distribution at peripheral regions of galaxy clusters.
It does change with the distance to the cluster center, but to the
first
appoximation the gas could be considered isothermal. Though the gas
temperature is possibly better defined observationally, it is more
difficult to make an estimate of the temperature in simulations like
ours without additional assumptions. We suppose that the gas
temperature is proportional to the velocity
dispersion of the dark matter in a cluster: $T\propto v^2$. The
relation can  be normalized to produce ``right'' temperature for the
Coma
cluster. If we take $v_{\rm coma}=1010$\kms (e.g. Zabludoff \etal 1990)
and $T_{\rm coma}=8.2{\rm keV}$ (Henry \& Arnaud 1991) then
$ T({\rm keV}) =(v/v_*)^2$, $v_*=350 {\rm kms}^{-1}$.
This is the relation, which we adopted for our analysis. The scale
$v_*$ is uncertain at best within 20 percent. In
Figure~3 we present
$dn/dT$ for galaxy clusters in the CHDM model.
The full curve shows the results for the model with the gas
temperature being estimated using the rms
velocity  found for the dark matter within radius $1.5\Mpch$.
Triangles in the plot indicate
the temperature distribution function of Abell clusters obtained
 by Henry \& Arnaud (1991) and the results of Edge \etal (1990) are
shown as squares. Bartlett \& Silk (1993) used the Press-Schechter
approximation ($\delta_c=1.68$) and the relation $T \propto M^{2/3}$
to estimate the temperature distribution function. They predict
significantly more high temperature clusters for the CHDM model
than the present study but the difference could be explained by
a scaling down the temperature estimated by Bartlett \& Silk by a
factor  $T/1.4$, which is not that large taking into account
the uncertainties in both the approximation for the number of clusters
and in the  T - M relation.
\bigskip
\centerline{\bf 6. CORRELATION FUNCTIONS}

In Figure~4 the correlation function of Abell
clusters of richness ${\cal R} \ge 0$ is shown as big circles.
We show the data of Postman \etal (1992) scaled to $\Omega=1$
with the addition of more accurate redshifts (Postman, 1993) and a
more accurate estimator for the correlation function: (DD-2DR+RR)/RR
(Landy \& Szalay 1993).
The full curve shows predictions of the CHDM model. For
the mass limit we chose $M > 2.5\times 10^{14}h^{-1}\Msun$. This is a
compromise between the number density of the Abell clusters and mass
of the clusters estimated by Bahcall \& Cen. In the
simulations there are slightly more clusters above the threshold as
compared
with the expected number of Abell clusters (148 versus 110) and the
threshold mass
of the ``clusters'' is slightly higher than the mass threshold
estimated by Bahcall \& Cen ($M > 1.8\times 10^{14}h^{-1}\Msun$).
The dashed line  in the plot shows the power law:
$ \xi_{\rm cc}(r) =(r/20h^{-1}{\rm Mpc})^{-1.8}, $
which at scales less than $(20-30)h^{-1}\Mpc$ gives reasonably good
fit for both the observational points and theoretical predictions. At
larger radii the correlation function falls below the power law and
becomes negative at $r > 50h^{-1}\Mpc$. The dashed line on the linear
part of the Figure presents the correlation function of the dark matter
estimated by the linear theory and scaled up by the factor $b^2=6^2$ to
match the correlation function at $40h^{-1}\Mpc$.

In Figure~5 we show the correlation function of APM clusters
(circles; Dalton \etal 1992) and predictions from the numerical
simulations
(the full curve). The mass limit for the APM-style clusters in the
model was $M > 1.05\times 10^{14}h^{-1}\Msun$ and the number of the
clusters in the simulation box was 203 and 206 for the two runs.
Error bars shown in  figures 4 and 5 are computed using poissonian
errors $(1+\xi) / \sqrt{\rm N_{pairs}}$.
\bigskip
\centerline{\bf 7. DISCUSSION}
\noindent
Four recent papers have determined that the cluster correlation
function becomes negative on large scales (Postman \etal 1992 and
Postman 1993,
Peacock and West, 1992, Dalton \etal 1992, Scaramella \etal 1993).
The zero point we infer from these four studies is
$\r0 =(50\pm10)\Mpch$.
This seems to be a robust result as the correlation functions have
been computed using different samples and different methods.
It is
clear from previous studies that the CDM model cannot reproduce
the cluster-cluster correlation function (Bahcall and Cen, 1992,
Olivier \etal 1993) because it has a zero point of $\r0 =33\Mpch$.
For the CDM model with a cosmological constant (Kofman \etal 1993)
the zero point occurs at $r_{0,CDM+\Lambda}=16.5(\Omega h^2)^{-1}$Mpc.
The observed limits on the zero point put severe constraints on this
model. The observed zero point lies in the range $40-60\Mpch$
implying the $\Omega h$ lies in the range 0.27--0.41. Thus models
with $h<1$ and $\Omega < 0.25$ are in conflict with the observations.
Assuming the age of the universe to be larger than 15~Gyr (to get a
benefit from introducing the cosmological constant) rejects
all models with $\Omega < 0.5$ or $h > 0.55$. For the tilted
CDM model the constraint on the zero point implies that the
large scale slope of the power spectrum should be in the range
$h=0.6-0.8$. The CDM + $\lambda$ model
and low $\Omega$ models claimed by Bahcall and Cen (1992) to fit the
data at $r<25\Mpch$ but do not account for the key observation of the
cluster zero point (i.e. their model correlation functions go negative
only beyond $r=100\Mpch$ in conflict with the observations). As shown
in
figures 4 and 5
the CHDM model provides an excellent fit to the APM and Abell
cluster correlation functions over the range $5 < r < 100\Mpch$
for which the function has been determined.

Note that two systematic errors can affect the determination of
the zero point. Firstly for a finite sample the correlation function
will be negative on large scales. The amplitude of this effect
is $-n_c/N$ where $n_c$ is the number of clusters per clump and N
is the total number. For the Abell sample $n_c$ is $\sim 2$
and the total number N is $\sim 200$. This effect alone would produce
a negative value of $\xi_{cc}$ on large scales of
$\sim -0.01$. Note however that observed and predicted negative amplitude
is $\sim -0.05$ considerably larger than this effect.

Secondly, the finite box size or observtional volume does not have long
waves which should have the efffect of reducing the zero-point.
The amplitude of this effect depends on the spectrum.
For the CHDM model and a 400Mpc box we estimate this to be a 10-20\% effect.
Results of our numerical simulations indicate that small
nonlinear effects on 50Mpc scales move the zero point up
by almost the same amount, thus basically compensating
the effect of the finite box size.

We thus emphasize that
it is now necessary to go beyond the power law representation of
$\xi(r)$ since this is no longer a good fit to the available data.
\bigskip
We thank Marc Postman
for sending us his correlation function via electronic mail.
\medskip
\centerline{FIGURE CAPTIONS}
\medskip
\item{Figure 1.}Number density of Abell clusters in the CHDM model as a
function of redshift (full curves and filled circles).
The dashed curves show the PS results for $\delta_c=1.45$.

\item{Figure 2.}The cluster mass function. The solid curve shows
the mass function predicted by the CHDM model. The dot-dashed curve
is the mass function obtained by the Press-Schechter approximation
with the parameter $\delta_c=1.5$.
The short- dashed curve shows  the mass function of Abell clusters
estimated by
Bahcall \& Cen (1993). The long-dashed curve provides a
very accurate fit to the mass function in the CHDM model. It is the
same fit as found by Bahcall \& Cen, but the mass scale is higher by
the factor 1.6.
\smallskip
\item {Figure 3.}The X-ray temperature distribution function.
The full curve shows the results for the rms
velocity  found for the dark matter within radius $1.5\Mpch$.
Triangles in the plot indicate
the temperature distribution function of Abell clusters obtained
 by Henry \& Arnaud (1991) and the results of Edge \etal (1990) are
shown as squares.
\smallskip
\item {Figure 4.}The Abell cluster correlation function.
The correlation function of Abell
clusters of richness R $\ge 0$ is shown as big circles (Postman
\etal 1992, Postman 1993).
The full curve shows predictions of the CHDM model.
The dashed line  in the plot shows  the standard power law correlation
function with a correlation length of $20\Mpch$.
The dashed line on the linear part of the plot presents the
correlation function of the dark matter predicted  by the linear theory.

\item {Figure 5.}The APM cluster correlation function.
The correlation function of APM clusters (Dalton \etal) is shown as
circles and the full curve shows predictions from the numerical simulations.
The mass limit for the APM-style clusters in the
model was $M > 1.05\times 10^{14}h^{-1}\Msun$ and the number of the
clusters in the simulation box was 203 and 206 for the two runs.
\smallskip
\vfill\eject
\frenchspacing
\centerline{\bf REFERENCES}
\vskip -0.5cm
\noindent
\def\hang{\par\hangindent=\parindent\noindent}
\def\author#1{\setbox0=\hbox{#1}\leavevmode\copy0}
\def\same{\leavevmode\hbox to \wd0{\leaders\hrule height
3.5pt depth-3pt\hfill}, }
\hang {Bahcall, N.A., and Soneira, R., 1983, ApJ, 270, 20}
\hang {Bahcall, N.A., and Cen, R., 1993, ApJL, 398, L81}
\hang {Bartlett, J.G., \& Silk, J., 1993 ApJL, L45}
\hang {Biviano, A., Girardi, M., Giuricin, G., Mardirossian, F. \&
Mezzetti M., 1993, ApJ, 000, 000}
\hang {Dalton, G.B., Efstathiou, G., Maddox, S.J., \& Sutherland, W., 1992,
ApJ, 390, L1}
\hang {Davis, M. and Peebles, P.J.E., 1983, ApJ, 267, 465}
\hang {Hockney, R.W., \& Eastwood, J.W., 1981, {\it Numerical
simulations using particles (New York: McGraw-Hill)}}
\hang {Jing, Y.P., Mo, H.J., Borner, G., \& Fang, L.Z., 1993, ApJ, 411, 450}
\hang {Kates, R.E., Kotok, E.V., \& Klypin, A.A., 1991, A\&A, 243, 295}
\hang {Klypin A.A. and Kopylov, A.I., 1983, SvA 9, L41}
\hang {Klypin, A. Holtzman, J. Primack, J. \& Reg\"os, E, 1993, ApJ, in press}
\hang {Kofman, L., Gnedin, N. \& Bahcall, N., 1993, preprint}
\hang {Lahav, O., Edge, A.C., Fabian, A.C, \& Putney, A., 1989, MNRAS, 238,
881}
\hang {Landy, S.D., Szalay, A.S., 1993, ApJ, 412, 64}
\hang {Nichol, R.C., Collins, C.A., Guzzo, L., \& Lumsden, S.L., 1992,
MNRAS, 255, 21}
\hang {Olivier, S.S., Primack, J.R., Blumenthal, G.R., \& Dekel, A., 1993,
preprint}
\hang {Peacock, J.A., \& West, M.J., 1992, MNRAS, 259, 494}
\hang {Postman, M., Huchra, J.P. \& Geller, M.J., 1992, ApJ, 384, 404}
\hang {Postman, M., 1993, private communication}
\hang {Scaramella, R., Zamorani, G. and Vettolani, G., 1993, preprint}
\hang {Sutherland, W.J., 1988, MNRAS, 234, 159}
\hang {Sutherland, W.J. \& Efstathiou, G., 1991 MNRAS, 248, 159}
\hang {West, M.J., \& van den Bergh, S., 1991, ApJ, 373, 1}
\hang {White, S.D.M., Navarro, J.F., Evrard, A.E., \& Frenk, C.S., 1993,
preprint}
\bye